\documentclass[aps,amsmath,prl,a4paper,floatfix,twocolumn]{revtex4}
\usepackage{graphicx}
\usepackage{amsmath}
\usepackage{amsfonts}
\usepackage{amssymb}

\newcommand{\beq}{\begin{equation}}
\newcommand{\eneq}{\end{equation}}
\newcommand{\bea}{\begin{eqnarray}}
\newcommand{\enea}{\end{eqnarray}}

\begin{document}
\title{Advantages of using YBCO-Nanowire-YBCO heterostructures in the search for Majorana Fermions}

\author{P. Lucignano$^{1,2}$} \author{A. Mezzacapo$^3$}  \author{F.Tafuri$^{4,5}$}\author{A. Tagliacozzo$^{2,5}$} 

\affiliation{$^1$ CNR-ISC, via Fosso del Cavaliere 100,  I-00133 Roma, Italy}
\affiliation{$^2$ Dipartimento di Scienze Fisiche, Universit\`a di Napoli ``Federico II'', Monte S.Angelo, I-80126 Napoli, Italy}
\affiliation{$^3$Departamento de Qu\'imica F\'isica  Universidad del Pa\'is Vasco - Euskal Herriko Unibertsitatea, Apdo. 644, 48080 Bilbao, Spain}    
\affiliation{$^4$ Dipartimento Ingegneria dell'Informazione, Seconda Universit\`a  di Napoli, I-81031Aversa (CE), Italy}
\affiliation{$^5$ CNR-SPIN, Monte S.Angelo -- via Cinthia,  I-80126 Napoli, Italy}

\date{\today}

\begin{abstract}
We propose an alternative platform to observe Majorana bound states in solid state systems. High critical temperature cuprate  superconductors can induce superconductivity, by proximity effect, in quasi one dimensional nanowires with strong spin orbit coupling.  They favor a wider and more robust range of Êconditions to stabilize Majorana fermions due to the large gap values, and offer novel functionalities in the design of the experiments determined by different dispersion for Andreev bound states as a function of the phase difference.
\end{abstract}

\maketitle

Recently there is an increasing interest in topological quantum computation based on Majorana Bound States (MBS's) \cite{Alicea:2012, Kitaev:2003}.
Majorana Fermions  have been predicted in a wide class of low-dimensional solid state devices. Many of these proposals make use of quasi one dimensional superconductors in contact with topological insulators \cite{Teo:2010} or quasi one-dimensional materials with strong spin orbit interactions  \cite{Lutchyn:2010,Oreg:2010,Duckheim:2011,Weng:2011}. Also helical magnets \cite{Kjaergaard:2011} and other materials \cite{Alicea:2010,Fu:2009,Akhmerov:2009,Tanaka:2009,Linder:2010} are considered.
In this paper  we propose a quite distinctive heterostructure to observe topologically protected  MBS's in a solid state device.
Our work rests on the physics  of S/R/S  hybrid structures in which "R"  is a quasi one dimensional semiconductor nanowire  (NW) with strong Rashba spin orbit coupling (e.g. InAs or InSb) electrically connected to two conventional low $T_c$ superconductor leads ( "S"  ) \cite{Lutchyn:2010,Oreg:2010}.
Superconductivity is induced  in the spin-orbit coupled semiconductor by proximity effect due to the superconducting electrodes. The coexistence of superconductivity and  spin-orbit coupling is a key ingredient for the existence of MBS's at the interfaces between the R region and the superconducting S regions.

However, despite the considerable theoretical and experimental \cite{Reich:2012} efforts, some challenges still remain before a real device allowing isolation and manipulation of MBS's  in such geometry can be realized. In particular the difficulties  of tuning the chemical potential of the semiconductor region  $\mu $, controlling  the disorder on the bulk gap as well as  optimizing the coupling between the different materials  \cite{Potter:2011,Lutchyn:2011,Brouwer:2011} make the realization of such devices extremely difficult.




All schemes proposed up to now to generate MBS's substantially use conventional s-wave superconductors  to induce superconductivity  and a gap $\Delta $  in the R nanowire\cite{Alicea:2012}. The role of superconducting pairing is  to relax number conservation, thus allowing for the  mixing of particle and hole degrees of freedom. Zeeman spin splitting is required to halve the number of degrees of freedom at low energies, thus generating  the elusive neutral (Majorana) excitation.  A simple criterion to induce MBS's  at S/R interfaces is given in terms of the applied magnetic field  $B_x$ oriented along the wire, $\mu $ and $\Delta $. The inequality to be satisfied can be stated as: $ B_x^2  > \mu ^2 +\Delta ^2 $\cite{Lutchyn:2010}.
Low critical  magnetic fields ($H_c$) and  low gap values characteristic of conventional low Tc superconductors substantially define the limits of the nominal range of dynamical parameters required to observe MBS's. Not only do  $H_c$ and $\Delta$  enter  into the criteria to stabilize MBS's, but they also endanger the feasibility of the experiment in case high magnetic fields are required. High critical temperature superconductors (HTS) may favor a completely different approach to experiments on MBS's, since HTS plaquettes/contacts (even a few micron square) sustain superconductivity up to a few tenths of Tesla and induce robust superconductivity in a wide range of barrier materials. When conventional low-Tc superconductors (e.g. Nb) are considered, the large  difference in the g-factors for Nb ($g_{Nb}\sim 1$)  and InAs ($g_{InAs}\sim35$) implies that the in-plane magnetic field $B\sim 0.1 T$ can open a sizable Zeeman gap in InAs ($V_x \le 1K$) without  substantially suppressing the superconductivity in Nb.  However  these conditions holds even more firmly in  YBCO contacts because  the YBCO gap is very stable  w.r.t. magnetic fields, despite a doubling of the $g-$factor ($g_{YBCO}\sim 2$).


 The induced gap in the NW  can be  considered of the order of the bare gap of the superconductor, projected along the wire direction, provided that  the radius of the wire is negligibly small with respect  to the coherence length $\xi$ of the superconducting material and that no sizeable barriers are present at the interfaces. 
As recently pointed out in Ref. \cite{Potter:2011}, the  interface tunnelling between  different materials renormalizes the induced gap to $\tilde \Delta_i = (1-Z)\Delta_i$ where $Z\sim(1+\pi \rho_0 |V_{hop}|^2/\Delta_i)^{-1} $is the quasiparticle weight,  $V_{hop}$ mimicks the electron hopping between the superconductor and the NW  and $\rho_0$ is the density of states of the superconductor at the Fermi energy.  The better is the coupling with a larger $V_{hop}$, the smaller becomes  $Z$ and  the larger is the induced gap. $Z$ also renormalizes the whole NW Hamiltonian $H_{NW} \rightarrow \tilde{H}_{NW}=Z H_{NW}$ which means that, by the same token, all the NW Hamiltonian parameters are effectively reduced when $V_{hop}$  increases.  When taking the renormalization  into account in the model that we discuss below,  the criterion for the appearance of the topologically non trivial phase given by Eq.\ref{bdominated}  becomes
\beq
Z ^2(B^2 - \mu^2 ) > (1-Z)^2\times  \max(|\Delta_L|^2,|\Delta_R|^2) \label{bdominated2}\:.
\eneq
This renormalization effect requires caution in the nanostrucure design and, interestingly enough, it can be  fruitfully  exploited in the case of HTS  proximity.  A convenient tradeoff can be found by  accepting a rather poor intermaterial coupling $V_{hop}$ , due to  the very large bare  superconducting gap along the lobe direction, which is almost one order of magnitude larger than in conventional low Tc superconductors. The InAs nanowire mostly rules the scaling of the proximity effect \cite{Kirtley:2005, tafuri_rop} once good interface conditions  are guaranteed between the HTS material and the InAs nanowire\cite{Montemurropreprint}.
The magnetic field can be very high with negligible effects both on the superconducting properties of the HTS electrode and on the interface transparency. 

Here we focus on other functionalities of HTS hybrid devices which are offered by an anisotropic  d-wave order parameter symmetry\cite{Tanaka:1995}. In d-wave systems lobes  in the excitation gap of amplitude 20 meV coexist with nodes, while in conventional s-wave superconductors the gap value is about or less than 1 meV and uniform in all directions.   In HTS contacts, the crystal axes' orientations with respect to the nanowire can be chosen in order to maximize the proximity induced $\Delta $. Different crystal orientations can be currently achieved by bicrystal or biepitaxial techniques \cite{tafuri_rop}. 

\begin{figure}[htbp]
\begin{center}
\includegraphics[width=\linewidth]{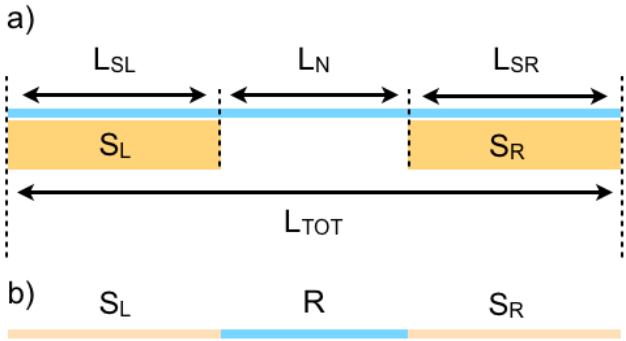}
\caption{a) Side view of the Superconductor-InAs nanowire-Superconductor heterostructure. b) Scheme of the structure used for the effective one-dimensional model.}
\label{fig1}
\end{center}
\end{figure}

\begin{figure}[htbp]
\begin{center}
\includegraphics[width=\linewidth]{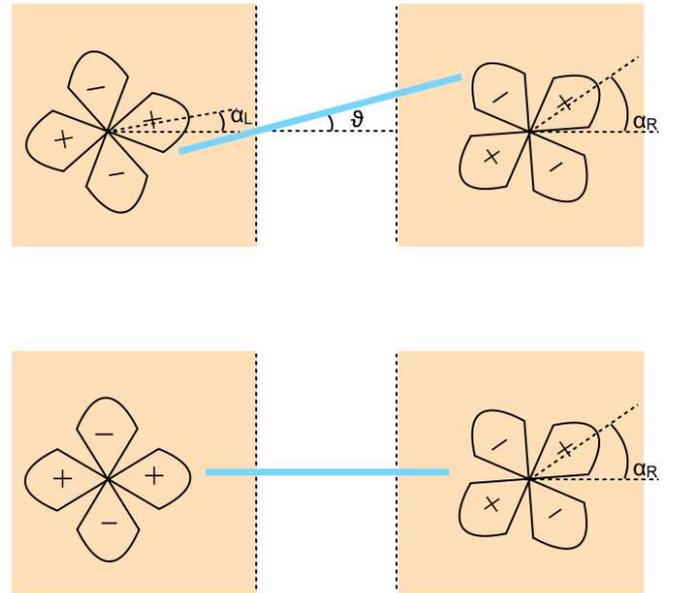}
\caption{The top view  sketch for different geometries. Configurations of the order parameter are determined by a suitable orientation of the electrodes and of the nanowire.}
\label{fig2}
\end{center}
\end{figure}

For sake of simplicity, we model the system as an effective  one-dimensional device 
composed by  the NW of length $L_N$ and two superconducting regions (see Fig.\ref{fig1}b), whose effective gaps differ not only in phase but also in their modulus, depending on the relative crystal orientation  (see Fig.\ref{fig2}).
Effective Hamiltonian parameters, including the interface renormalization, will be disregarded, for the moment, and reintroduced in a second step.

 In the superconducting regions of the nanowire,  spin-orbit interaction and superconductivity coexist{\it}. We assume that $L_N<<\xi  << L_{TOT}$  to have penetration of superconductivity in the whole nanowire.
 
\begin{widetext}
A Bogoliubov-De Gennes  mean field Hamiltonian fully accounts for superconductivity induced in the normal material by proximity effect:
\bea
{\cal{H}}_S = \left ( H_0 -\mu \: N \right )+ \int_{-\infty}^{\infty}\: d x \left(\Delta(x) \psi^\dagger _\uparrow (x)  \psi^\dagger _\downarrow (x) + h.c. \right) \:,
\nonumber\\
H_0  -\mu\:  N = \int_{-\infty}^{\infty}\: d x \: \psi^\dagger _\alpha (x) \left(\left(- \frac{\partial_x^2}{2m^*} -\mu\right)I_2 + i\eta \:  \sigma_y \partial_x +B_x \sigma_x \right)_{\alpha \beta}\psi_{\beta} (x)
\enea
where $x$ is the coordinate along the wire and  $\alpha,\beta = \uparrow,\downarrow$ denote the two components of the electronic fermionic fields. $m^*$ and $\eta $ are the effective mass  and the Rashba spin orbit coupling strength, respectively. $B_x= g \mu_B B/2$ is the effective  Zeeman spin splitting energy. It is assumed that the magnetic field, chosen in the direction of the wire, does not induce any undesired  orbital effect .
The  $ d_{x^2-y^2} $ superconductivity  pairing  is modeled as\cite{Tsuei:2000}:  
\beq
\Delta(x) = 
\left\{\begin{array}{llr}
\Delta_L& = \Delta_0 \: cos(2(\vartheta-\alpha_L)) &\;\;\;for\;\; x<-L_N/2\:, \\
0& & \;\;\;for\;\;  -L_N/2\le x  \le L_N/2\:, \\
\Delta_R& =\Delta_0 \: e^{-i \phi}\: cos(2(\vartheta-\alpha_R))& \;\;\;for\;\; x>L_N/2\:,
\end{array}
\right.
\eneq
\end{widetext}
Angles $\alpha _{R,L} ,\vartheta $ are defined in Fig.\ref{fig2}. $\phi $ is an $U(1)$  phase difference  across the junction. Let us choose  $\vartheta$  zero. Depending on the relative orientation of the order parameters in the L,R regions with respect to the nanowire, a wealth of possibilities occur.  For an effectively one-dimensional wire  we can  set  $\alpha_L=0$ with no loss of generality.  By rotating $\alpha_R$ from $0$ to $\pi/2$ we can continuously explore all the configurations from lobe-lobe (+/+) to lobe-antilobe(+/-). Nodal configurations are not interesting here, as we need large superconducting gaps. As we are searching for MBS's, we will choose  only a few angle configurations to demonstrate the main concepts.

The Hamiltonian operator in  $ {\cal{H}}_S $ can be recast in the compact form, in the basis $\hat{\psi}(x) = [\psi_\uparrow (x),\psi_\downarrow(x),\psi _\downarrow\: ^\dagger (x),- \psi _\uparrow\: ^\dagger (x)]$:
\beq
\tilde {H}_S/\eta  = \left[ \left (- \frac{1}{2} \: \partial_x^2-\mu \right )  \sigma_0 +i \partial_x \sigma_y \right]\: \tau_z + B_x \sigma_x \tau_0 + \Delta(x)\tau_x \: . 
\label{hamiltonian}
\eneq
It is a  tensor product of matrices $\tau_i \times \sigma_j$ with $\{i,j\} \in \{0,1,2,3\}$, where  $\tau_i$ and $\sigma_i$ are the usual Pauli matrices for $i\neq0$ and  the $I_2$ identity matrix  for $i=0$. They refer to the Nambu and spin  degrees of freedom, respectively.
 A new space scale  $x \to \eta \:  m^*\: x$ has been introduced, as well as  energy scale:  $\mu, B_x ,\Delta  \to  \mu , B_x, 
 \Delta \: /(m^*\: \eta^2)$.
 
 The Hamiltonian of Eq.\ref{hamiltonian} has a topologically non-trivial phase whose boundary states are Majorana Fermions  \cite{Lutchyn:2010,Oreg:2010}, provided 
 \beq
B_x^2 - \mu^2  > \max(|\Delta_L|^2,|\Delta_R|^2) \:,
\label{bdominated}
\eneq
where the hamiltonian parameters used here, effectively include the quasiparticle renormalization  weight $Z$.
The  topologically trivial phase,  is adiabatically deformable to the usual Andreev physics \cite{Blonder:1982}. We calculate numerically the low lying part of the energy spectrum by matching  the eigenfunctions. In order to simplify calculations, we take the limit $L_N\sim 0$ by matching  the wavefunction and its derivative at $x=0$.  As shown in \cite{Lutchyn:2010}, this assumption does not alter the generality of our results as interaction terms among Majorana end states are neglected in our approach. The effects of a finite size wire are shown for example in Ref.\cite{Pikulin:2012}.

\begin{figure}[htbp]
\begin{center}
\includegraphics[width=\linewidth]{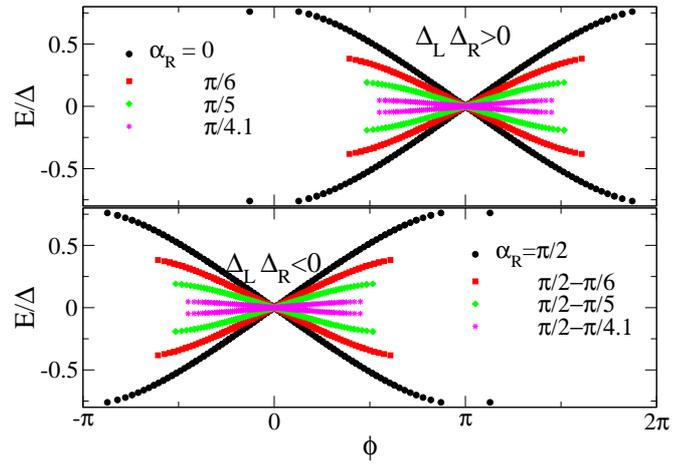}
\caption{Energy spectrum of zero energy Majorana bound states in the case of equal (opposite) sign gaps (in top and bottom panel respectively).}
\label{fig3}
\end{center}
\end{figure}

In Fig. \ref{fig3}  the dispersion relation of MBS's is shown as a function of the phase difference, $\phi$, between the superconducting pads. For $\alpha_R <\pi/4$ the Andreev levels show a single crossing at $\phi=\pi$. The odd number of crossings in the Andreev spectrum is the characteristic signature of the topological non trivial phase, consistent with that found with conventional s-wave superconductivity.  However, the Andreev spectrum  shows an unexpected behavior when $\alpha_R >\pi/4$ i.e. when the effective induced gaps have opposite signs. 
In this case, the crossing, which features the zero energy MBS, is still present, but located at  $\phi = 0$. This is  specific of the d-wave order parameter. When $\alpha_R < \pi/4$ the gaps $\Delta_L$ and $\Delta_R$ have the same sign. Therefore, a phase difference of $\pi$ between the two order parameters  is required, in order to have an inversion of the  sign of the gap between the two regions $S_1$ and $S_2$. Provided that  the appropriate condition for the parameters is met, the sign inversion,  irrespective of the relative strength of the two gaps (and of the actual value of $\alpha _R$), enforces   the crossing to be localized at $\phi=0$, and the Majorana excitation with it.
 Together with this change, the shape of the dispersion relation changes by changing $\alpha _R$,  with an increase of  the current  $I(\phi) = \partial E(\phi) / \partial \phi $ up to a maximum, when  the gaps reach their  maximum at  $\alpha_R=0$ or $\pi/2$. The crossing only appears at $\phi=0 ,\pi$, because only at these points  the Hamiltonian is real. 
Moreover, depending on the crystal relative arrangements we can have a different dispersion for Andreev Bound states as a function of the phase difference $\phi$.  In both cases a single crossing at zero energy appears, which reveals the presence of the MBS, at $\phi=0$ or $\pi$ depending on the sign of the product $\Delta_L \Delta_R$. 

At present, the race to detect  signatures of the elusive Majorana Fermions in an S/NW/S structure is quite exciting \cite{Goldhabergordon:2012, Kouwenhoven:2012, Heiblum:2012,Rokhinson:2012}. A system exploiting d-wave electrodes, as the one proposed in this work, can inspire hallmark experiments in the search  for
Majorana excitations.

Despite the fact that the magnetic field should dominate over the superconductivity, still, a sizable superconducting gap is needed, as the smaller energy between $V_z$ and $\Delta $ sets the minimum  energy sufficient  to wash out the topological protection of the Majorana excitation. In this respect, HTS's appear to offer more chances in stabilizing MBS's. 
Question arises whether nodal
quasiparticles in the d-wave \cite{Tsuei:2000} topologically trivial superconductor could
affect their stability. As the MBS's imply strong non
local correlations, one is inclined to conclude that local nodal
quasiparticle should be inefficient in producing a decay of the Majorana
zero energy excitation. Besides, nodal quasiparticles are strictly at zero
energy if travelling along given directions in an uniform system d-wave ordering. The
presence of the Josephson barrier, inhomogeinity or  confined geometry,
should move those states to finite energy.   In the
different context of  YBCO grain boundary Josephson Junctions\cite{Tagliacozzo:2009,Lucignano:2010}, we have experienced  a long lasting quantum coherence of antinodal quasiparticles,
while an efficient relaxation mechanism could have been the production of
nodal quasiparticles\cite{Lucignano:2010,Gedik:2003}. This is  a conforting piece of evidence, but, of
course, not a proof, though.

 A d-wave induced superconductivity encompasses a wider range  of opportunities to discriminate the presence of the MBS.  Andreev states induced by d-wave pairing are strongly sensitive to  the geometry of the device.  The characteristic increase of the $I_c$ at the lower temperatures, used as a benchmark for the existence of the Andreev midgap  state in HTS junctions\cite{Tanaka:1995}, is strongly suppressed when  the width of the junction is reduced toward  a quasi 1D device.  However, in our devices, an anomalous increase of $I_c$ at low temperatures would persist in the one-dimensional limit and would be even  sharper, the lower the barrier transparency is. This would unambiguously   signals feature that can be only correlated to the presence of Majorana fermions \cite{supp1,Zazunov:2012,Ioselevich:2011}. Prototype structures using  HTS electrodes are being already  tested \cite{Montemurropreprint}.

Moreover, it is a distinctive property of ring structures with appropriate multi crystal arrangements  to entail  frustrated d-wave pairing ordering with   trapped fractional fluxes in the ground state \cite{Tsuei:2000}. The possibility  highlighted in this work, to have MBS localized at $0-$  and $\pi-$ junctions, depending on the phase configuration, can be exploited in the design of  quantum coherent, topologically protected devices, which go beyond the simple experimental confirmation of this amazing new physics, to enter  the field of applications.  

We envisage   the possibility of  engineering quasi-degenerate odd fermionic parity  states, using a mesoscopic, charge isolated island, formed by a d-wave tricrystal \cite{supp1} topologically protected with respect to the  excitations.

 
\begin{acknowledgments}
We acknowledge important discussions with D. Bercioux, P. Brouwer, M. Cuoco, P. Gentile, D. Urban, F. von Oppen. P.L. acknowledges F. Dalton for a critical proofreading of the manuscript. Financial support from FP7/2007-2013 under the grant N. 264098 - MAMA (Multifunctioned Advanced Materials and Nanoscale Phenomena),   MIUR-Italy by Prin-project 2009 "Nanowire high critical temperature superconductor field-effect devices" , and European "SOLID" Project are gratefully acknowledged.
\end{acknowledgments}

\bibliographystyle{apsrev}


\end{document}